\documentclass[italian,english]{article}
\usepackage[latin9]{inputenc}
\usepackage{graphicx}
\usepackage{amssymb}

\newcommand{\lyxaddress}[1]{
\par {\raggedright #1
\vspace{1.4em}
\noindent\par}
}

\usepackage{babel}

\begin{document}

\title{\textbf{Interferometer Response to Scalar Gravitational Waves}}

\author{\textbf{C. Corda$^{1}$, S. A. Ali$^{2}$ and C. Cafaro$^{3}$}}

\maketitle

\lyxaddress{\begin{center}
\textbf{$^{1}$}\emph{Institute for Basic Research, P. O. Box 1577,
Palm Harbor, FL 34682, USA and Associazione Galileo Galilei, Via Pier
Cironi 16 - 59100 Prato, ITALY}
\par\end{center}}

\lyxaddress{\begin{center}
\textbf{$^{2}$}\emph{Department of Physics, State University of
New York at Albany, 1400 Washington Avenue, Albany, NY 12222, USA
and Department of Arts and Sciences, Albany College of Pharmacy and
Health Sciences, Albany, NY 12208, USA} 
\par\end{center}}

\lyxaddress{\begin{center}
\textbf{$^{3}$}\emph{Dipartimento di Fisica, Università di Camerino,
Via Madonna delle Carceri, 9 - Camerino, ITALY}
\par\end{center}}

\lyxaddress{\begin{center}
\textit{\emph{E-mail addresses:}}\textit{ }\textbf{\textit{$^{1}$}}\textit{cordac.galilei@gmail.com;
}\textbf{$^{2}$}\textit{alis@alum.rpi.edu; }\textbf{\textit{$^{3}$}}\textit{carlocafaro2000@yahoo.it}
\par\end{center}}
\begin{abstract}
It was recently suggested that the \emph{magnetic} component of Gravitational
Waves (GWs) is relevant in the evaluation of frequency response functions
of gravitational interferometers. In this paper we extend the analysis
to the \emph{magnetic} component of the scalar mode of GWs which arise
from scalar-tensor gravity theory. In the low-frequency approximation,
the response function of ground-based interferometers is calculated.
The angular dependence of the \emph{electric} and \emph{magnetic}
contributions to the response function is discussed. Finally, for
an arbitrary frequency range, the proper distance between two test
masses is calculated and its usefulness in the high-frequency limit
for space-based interferometers is briefly considered. 
\end{abstract}

\lyxaddress{\textit{Keywords}: Scalar Gravitational Waves; interferometers; \emph{magnetic}
components.}

\lyxaddress{PACS numbers: 04.80.Nn, 04.80.-y, 04.25.Nx}

\section{Introduction}

$\mbox{ }\mbox{ \mbox{ \mbox{ }}}$The study of the so-called \emph{magnetic}
component of Gravitational Waves (GWs) is an active field of research.
In a series of recent works, the significance of these \emph{magnetic}
components to the total frequency response function of omnidirectional
gravitational wave interferometers was emphasized \cite{key-1}-\cite{key-4}.
There are currently a number of gravitational wave detectors at various
stages of development worldwide. While several such projects are in
their design phase, others like the VIRGO detector (Cascina, Italy
\cite{key-5,key-6}) are already under construction. Yet other projects,
such as the GEO 600 (Hannover, Germany \cite{key-7,key-8}) the two
LIGO detectors (Washington and Louisiana, USA \cite{key-9,key-10})
and the TAMA 300 detector (Tokyo, Japan \cite{key-11,key-12}) are
functional and already taking scientific data. In view of such enormous
international effort, it is prudent to carry out a thorough analysis
of the expected frequency response associated with gravitational waves
predicted by General Relativity and its various competing theoretical
models of which scalar-tensor gravity is one. Indeed, GW detectors
will prove invaluable in the confirmation (or contradiction) of the
physical consistency of the various theories of gravity \cite{key-13}-\cite{key-22}.
Baskaran and Grishchuk have recently discussed the existence and relevance
of the so-called \emph{magnetic} components of GWs, which must be
taken into account in the context of the total response functions
(angular patterns) of interferometers for GWs propagating from arbitrary
directions \cite{key-3}. In \cite{key-4}, more detailed angular
and frequency dependence of the \emph{magnetic} contribution to Interferometer
Response Functions was given with specific application to the parameters
of the LIGO and VIRGO interferometers. 

In this paper the analysis is extended to the \emph{magnetic} component
of the scalar mode of gravitational waves which arise from scalar-tensor
theories of gravity \cite{key-14,key-15,key-16,key-19,key-21,key-22}.
The angular dependence of the response function of interferometers
for this \emph{magnetic} component is given in the approximation of
wavelength much larger than the linear dimensions of the interferometer.
The results of this paper generalize the works of \cite{key-3,key-4}
where it was shown that the \emph{electric} and \emph{magnetic} contributions
are unambiguous in the long-wavelength approximation for ordinary
GWs arising from General Relativity. In the high frequency regime,
the division into \emph{electric} and \emph{magnetic} components becomes
ambiguous. For this reason, in order to calculate the response of
a GW interferometer in this regime, we shall use the exact (i.e.,
without low-frequency approximation) expressions for distance measurements. 

The paper is organized as follows. In Section 2, a review of GWs from
scalar-tensor theories of gravity is presented. An analysis of GWs
arising from scalar-tensor gravity theory in the frame of a local
observer is implemented in Section 3. In Section 4, we obtain the
distance function between the two test particles of the interferometer
by integrating the geodesic deviation equation. The variation of distances
between test masses and the response of interferometers are considered
in Section 5. In Section 6, we obtain the interferometer test mass
distance function in absence of low-frequency approximation. It is
then shown that the \emph{electric} and \emph{magnetic} contributions
of the distance function arise from the series expansion of this exact
result. Our conclusions are presented in Section 7.

\section{Scalar gravitational waves from scalar-tensor theories of gravity}

$\mbox{ }\mbox{ \mbox{ \mbox{ }}}$If the gravitational Lagrangian
is nonlinear in the curvature invariants, the corresponding Einstein
field equations have an order higher than second \cite{key-14,key-15,key-16,key-19,key-21,key-22}.
For this reason such theories are often called higher-order gravitational
theories. In the most general case, such higher order theories arise
from an action of form 

\begin{equation}
S=\int d^{4}x\sqrt{-g}[F(R,\square R,\square^{2}R,\square^{k}R,\phi)-\frac{\epsilon}{2}g^{\mu\nu}\phi_{;\mu}\phi_{;\nu}+\mathcal{L}_{m}],\label{eq: high-order}\end{equation}
where $F$ is an unspecified function of curvature invariants and
a scalar field $\phi$, $D_{\mu}\equiv\partial_{\mu}+\Gamma_{\mu}$
is the covariant derivative with respect to the Christoffel connection
coefficients and $\square\equiv g^{\mu\nu}\partial_{\mu}\partial_{\nu}$
is the d'Alembertian operator. The semicolon in $\phi_{;\mu}$ represents
$D_{\mu}$ -differentiation. The term $\mathcal{L}_{m}$ is the minimally
coupled ordinary matter contribution.

In scalar-tensor theories of gravity, both the metric tensor $g_{\mu\nu}$
and a fundamental scalar field $\phi$ are involved \cite{key-14,key-19}.
The action of scalar-tensor gravity theory can be recovered from (\ref{eq: high-order})
with the choice 

\begin{equation}
\begin{array}{ccc}
F(R,\phi)=f(\phi)R-V(\phi) & and & \epsilon=-1\end{array}.\label{eq: ST}\end{equation}
Note that in this article we work with geometrized units $G=1$, $c=1$
and $\hbar=1$. Considering the choice (\ref{eq: ST}), the most general
action of scalar-tensor gravity theory in four dimensions is given
by 

\begin{equation}
S=\int d^{4}x\sqrt{-g}[f(\phi)R+\frac{1}{2}g^{\mu\nu}\phi_{;\mu}\phi_{;\nu}-V(\phi)+\mathcal{L}_{m}].\label{eq: scalar-tensor}\end{equation}
By choosing

\begin{equation}
\begin{array}{ccc}
\varphi=f(\phi), & \omega(\varphi)=\frac{f(\phi)}{2f'(\phi)}, & W(\varphi)=V(\phi(\varphi)),\end{array}\label{eq: scelta}\end{equation}
eq. (\ref{eq: scalar-tensor}) reduces to

\begin{equation}
S=\int d^{4}x\sqrt{-g}[\varphi R-\frac{\omega(\varphi)}{\varphi}g^{\mu\nu}\varphi_{;\mu}\varphi_{;\nu}-W(\varphi)+\mathcal{L}_{m}],\label{eq: scalar-tensor2}\end{equation}

Action (\ref{eq: scalar-tensor2}) represents a generalization of
Brans-Dicke theory \cite{key-22}. Variation of action (\ref{eq: scalar-tensor2})
with respect to $g_{\mu\nu}$ and $\varphi$ results in the Einstein-like
equation \begin{equation}
G_{\mu\nu}=-\frac{4\pi\tilde{G}}{\varphi}T_{\mu\nu}^{(m)}+\frac{\omega(\varphi)}{\varphi^{2}}(\varphi_{;\mu}\varphi_{;\nu}-\frac{1}{2}g_{\mu\nu}g^{\alpha\beta}\varphi_{;\alpha}\varphi_{;\beta})+\frac{1}{\varphi}(\varphi_{;\mu\nu}-g_{\mu\nu}\square\varphi)+\frac{1}{2\varphi}g_{\mu\nu}W(\varphi)\label{eq: einstein-general}\end{equation}
and Klein-Gordon equation

\begin{equation}
\square\varphi=\frac{1}{2\omega(\varphi)+3}(-4\pi\tilde{G}T^{(m)}+2W(\varphi)+\varphi W'(\varphi)+\frac{d\omega(\varphi)}{d\varphi}g^{\mu\nu}\varphi_{;\mu}\varphi_{;\nu}\label{eq: KG}\end{equation}
respectively. The quantity $T_{\mu\nu}^{(m)}$ appearing in (6) is
the ordinary stress-energy tensor of matter while $\tilde{G}$ is
a dimensional, positive definite constant \cite{key-14,key-17,key-19}.
The Newton constant is replaced by the effective coupling

\begin{equation}
G_{eff}=-\frac{1}{2\varphi},\label{eq: newton eff}\end{equation}
which is, in general, different from $G$. General Relativity is obtained
when the scalar field coupling is 

\begin{equation}
\varphi=const.=-\frac{1}{2}.\label{eq: varphi}\end{equation}

When $\omega=\omega(\varphi)=const.$ in eqs. (\ref{eq: einstein-general})
and (\ref{eq: KG}) the field equations describe the so-called string-dilaton
gravity \cite{key-14,key-17,key-19}. Since we study gravitational
waves in this work, the linearized theory in vacuum ($T_{\mu\nu}^{(m)}=0$)
with a small perturbation  $h_{\mu\nu}$ of the background will be
analysed. 

It is assumed that the tensorial sector of the background is Minkowskian
$\eta_{\mu\nu}$ while the scalar sector is $\varphi=\varphi_{0}$
where $\varphi_{0}$ is assumed to be a minimum of $W,$ 

\begin{equation}
W\simeq\frac{1}{2}\alpha\delta\varphi^{2}\Rightarrow W'\simeq\alpha\delta\varphi.\label{eq: minimo}\end{equation}

For the perturbed background specified by

\begin{equation}
\begin{array}{ccc}
g_{\mu\nu}=\eta_{\mu\nu}+h_{\mu\nu}, &  & \varphi=\varphi_{0}+\delta\varphi.\end{array}\label{eq: linearizza}\end{equation}
the linearized field equations are given by \cite{key-14,key-19,key-23}

\begin{equation}
\begin{array}{ccc}
\widetilde{R}_{\mu\nu}-\frac{\widetilde{R}}{2}\eta_{\mu\nu}=-\partial_{\mu}\partial_{\nu}\Phi+\eta_{\mu\nu}\square\Phi, &  & {}\square\Phi=m^{2}\Phi,\end{array}\label{eq: linearizzate1}\end{equation}
with\begin{equation}
\begin{array}{ccc}
\Phi\equiv-\frac{\delta\varphi}{\varphi_{0}}, &  & m^{2}\equiv\frac{\alpha\varphi_{0}}{2\omega+3}\end{array}\label{eq: definizione}\end{equation}
 where $\widetilde{R}_{\mu\nu\rho\sigma}$, $\widetilde{R}_{\mu\nu}$
and $\widetilde{R}$ are the linearized versions of $R_{\mu\nu\rho\sigma}$,
$R_{\mu\nu}$ and $R$ respectively, and are moreover, computed to
first order in $h_{\mu\nu}$ and $\delta\varphi$.

The case in which $\omega=const.$ and $W=0$ in eqs. (\ref{eq: einstein-general})
and (\ref{eq: KG}) has been analysed in \cite{key-19} in a manner
that generalizes the canonical linearization of General Relativity
\cite{key-23}. In particular, the transverse-traceless (TT) gauge
has been generalized to scalar-tensor gravity obtaining the total
perturbation of a gravitational wave propagating in the $z-$direction
(in the TT gauge) as, \begin{equation}
h_{\mu\nu}(t+z)=A^{+}(t+z)e_{\mu\nu}^{(+)}+A^{\times}(t+z)e_{\mu\nu}^{(\times)}+\Phi(t+z)e_{\mu\nu}^{(s)}.\label{eq: perturbazione totale}\end{equation}
The term $A^{+}(t+z)e_{\mu\nu}^{(+)}+A^{\times}(t+z)e_{\mu\nu}^{(\times)}$
describes the two standard (i.e. tensorial) polarizations of gravitational
waves which arises from General Relativity in the TT gauge \cite{key-23},
while the term $\Phi(t+z)e_{\mu\nu}^{(s)}$ is the extension of the
TT gauge to the scalar case. For a purely scalar GW the metric perturbation
(\ref{eq: perturbazione totale}) reduces to 

\begin{equation}
h_{\mu\nu}=\Phi e_{\mu\nu}^{(s)},\label{eq: perturbazione scalare}\end{equation}
and the corresponding line element is \cite{key-19,key-21} \begin{equation}
ds^{2}=dt^{2}-dz^{2}-(1+\Phi)dx^{2}-(1+\Phi)dy^{2},\label{eq: metrica TT scalare}\end{equation}
with $\Phi=\Phi_{0}e^{i\omega(t+z)}.$

The worldline $x,y,z=const.$ is a timelike geodesic representing
the history of a free test mass (see the analogy with tensorial waves
in \cite{key-3,key-4,key-23}). In other words, in scalar-tensor theories
of gravity the scalar field generates a third tensorial polarization
for gravitational waves. This is because in equations (32) of \cite{key-19}
three different degrees of freedom are present, in contrast to standard
General Relativity where gravitational waves exhibit two degrees of
freedom. For details concerning the generation of this third tensorial
polarization, see Section 3 of \cite{key-19}.

\section{Analysis in a locally inertial reference frame }

$\mbox{ }\mbox{ \mbox{ \mbox{ }}}$In a laboratory environment on
Earth, a coordinate system (frame of local observer) in which space-time
is locally flat is typically used \cite{key-3,key-4,key-19,key-20,key-21,key-23}.
In this frame, the distance between any two points is given simply
by the difference in their coordinates (in the sense of Newtonian
physics). Moreover, in this frame, SGWs manifest themselves by exerting
tidal forces on masses (the mirror and the beam-splitter in the case
of an interferometer, see Figure 1). 

\begin{figure}
\includegraphics{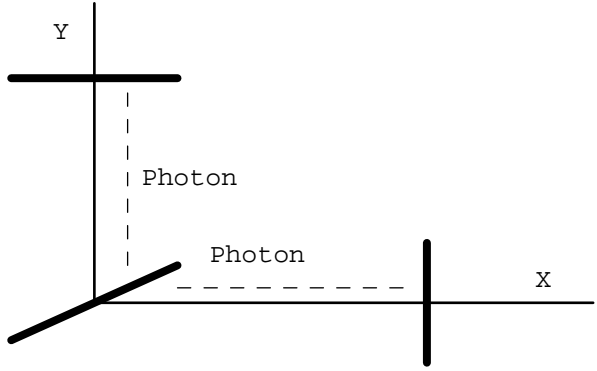}

\caption{Photons launched from the beam-splitter reflected back by the mirror.}

\end{figure}
A detailed analysis of the frame of the local observer is given in
ref. \cite{key-23}, Section 13.6. Here only the essential features
of this frame are emphasized. In particular, the coordinate $x_{0}$
is the proper time of the observer $O$; spatial axes are centred
in $O$; in the special case of zero acceleration and zero rotation,
the spatial coordinates $x_{j}$ are the proper distances along the
axes and the frame of the local observer reduces to a local Lorentz
frame. In this case the line element reads 

\begin{equation}
ds^{2}=-(dx^{0})^{2}+\delta_{ij}dx^{i}dx^{j}+O(x{}^{2}),\label{eq: metrica local lorentz}\end{equation}
where \begin{equation}
O(x{}^{2})\equiv\frac{1}{2}\left(\frac{\partial^{2}g_{\mu\nu}}{\partial x^{\sigma}\partial x^{\rho}}\right)_{P_{0}}(x^{\sigma}-x_{P_{0}}^{\sigma})(x^{\rho}-x_{P_{0}}^{\rho})dx^{\mu}dx^{\nu}\label{eq: zeri}\end{equation}
with $P_{0}$ being the point where the Lorentz frame is placed. The
effect of SGWs on test masses is described by the equation for geodesic
deviation in this frame 

\begin{equation}
\ddot{x}^{\mu}=-\widetilde{R}_{0\nu0}^{\mu}x^{\nu},\label{eq: deviazione geodetiche}\end{equation}
where $\widetilde{R}_{0\nu0}^{\mu}$ are the components of the linearized
Riemann tensor \cite{key-4,key-21,key-23}. 

Labelling the coordinates of the TT gauge as $t_{tt},x_{tt},y_{tt},z_{tt}$,
the coordinate transformation $x^{\alpha}=x^{\alpha}(x_{tt}^{\beta})$
from TT coordinates to the frame of the local observer is 

\begin{equation}
\begin{array}{c}
t=t_{tt}+\frac{1}{4}(x_{tt}^{2}-y_{tt}^{2})\dot{\Phi}\\
\\x=x_{tt}+\frac{1}{2}x_{tt}\Phi+\frac{1}{2}x_{tt}z_{tt}\dot{\Phi}\\
\\y=y_{tt}+\frac{1}{2}y_{tt}\Phi+\frac{1}{2}y_{tt}z_{tt}\dot{\Phi}\\
\\z=z_{tt}-\frac{1}{4}(x_{tt}^{2}-y_{tt}^{2})\dot{\Phi},\end{array}\label{eq: trasf. coord.}\end{equation}
where $\dot{\Phi}\equiv\frac{\partial\Phi}{\partial t}$ (see the
analogy with tensorial waves in \cite{key-3,key-4,key-24,key-25}).
The coefficients of this transformation (components of the metric
and its first time derivative) are taken along the central wordline
of the local observer \cite{key-3,key-24,key-25}. The linear and
quadratic terms, as powers of $x_{tt}^{\alpha}$, are unambiguously
determined by the conditions of the frame of the local observer \cite{key-3,key-4,key-24,key-25}. 

Considering a free mass with a timelike geodesic ($x=l_{1}$, $y=l_{2},$
$z=l_{3}$) \cite{key-3,key-4}, equations (\ref{eq: trasf. coord.})
define the motion of this mass with respect to the introduced frame
of the local observer, namely \begin{equation}
\begin{array}{c}
x(t)=l_{1}+\frac{1}{2}l_{1}\Phi(t)+\frac{1}{2}l_{1}l_{3}\dot{\Phi}(t)\\
\\y(t)=l_{2}+\frac{1}{2}l_{2}\Phi(t)+\frac{1}{2}l_{2}l_{3}\dot{\Phi}(t)\\
\\z(t)=l_{3}-\frac{1}{4}(l_{1}^{2}-l_{2}^{2})\dot{\Phi}(t).\end{array}\label{eq: Grishuk 0}\end{equation}
 In absence of GWs the position of the mass is $\vec{x}_{mass}=(l_{1},l_{2},l_{3}).$
The effect of the SGW is to drive the mass to have oscillations. Thus,
in general all three components of motion are present in (\ref{eq: Grishuk 0}).
Neglecting the terms with $\dot{\Phi}$ in (\ref{eq: Grishuk 0}),
the {}``traditional'' equations for the motion of the mass are obtained
\cite{key-19,key-21},\begin{equation}
\begin{array}{c}
x(t)=l_{1}+\frac{1}{2}l_{1}\Phi(t)\\
\\y(t)=l_{2}+\frac{1}{2}l_{2}\Phi(t)\\
\\z(t)=l_{3}.\end{array}\label{eq: traditional}\end{equation}
Clearly, this is analogous to the \emph{electric} component of motion
in electrodynamics \cite{key-3,key-4}, while equations\begin{equation}
\begin{array}{c}
x(t)=l_{1}+\frac{1}{2}l_{1}l_{3}\dot{\Phi}(t)\\
\\y(t)=l_{2}+\frac{1}{2}l_{2}l_{3}\dot{\Phi}(t)\\
\\z(t)=l_{3}-\frac{1}{4}(l_{1}^{2}-l_{2}^{2})\dot{\Phi}(t),\end{array}\label{eq: news}\end{equation}
are analogues of the \emph{magnetic} component of motion. One might
suspect the presence of this \emph{magnetic} component is a \textquotedblleft{}frame
artefact\textquotedblright{} due to transformation (\ref{eq: trasf. coord.}).
This issue will be addressed in the next Section where it will be
shown that (\ref{eq: news}) is obtained directly from the geodesic
deviation equation, verifying that the \emph{magnetic} components
have real physical significance. It is important to observe that the
\emph{magnetic} component become significant when the frequency of
the wave increases, but only in the low-frequency regime. This can
be understood directly from equations (\ref{eq: Grishuk 0}). In fact,
recalling that $\Phi=\Phi_{0}e^{i\omega(t+z))}$, (\ref{eq: Grishuk 0})
becomes\begin{equation}
\begin{array}{c}
x(t)=l_{1}+\frac{1}{2}l_{1}\Phi(t)+\frac{1}{2}l_{1}l_{3}\omega\Phi(\omega t-\frac{\pi}{2})\\
\\y(t)=l_{2}+\frac{1}{2}l_{2}\Phi(t)+\frac{1}{2}l_{2}l_{3}\omega\Phi(\omega t-\frac{\pi}{2})\\
\\z(t)=l_{3}-\frac{1}{4}(l_{1}^{2}-l_{2}^{2})\omega\Phi(\omega t-\frac{\pi}{2}).\end{array}\label{eq: Grishuk 01}\end{equation}
Thus, terms with $\dot{\Phi}$ in (\ref{eq: Grishuk 0}) can be neglected
only when the wavelength goes to infinity, while at high-frequencies
expansion terms proportional to $\omega l_{i}l_{j}$ $(i,j=1,2,3)$
corrections in (\ref{eq: Grishuk 01}) break down.

\section{Equations of motion from geodesic deviation}

$\mbox{ }\mbox{ \mbox{ \mbox{ }}}$ In this Section we consider the
geodesic deviation extended to second order approximation from which
it is shown that the \emph{magnetic} component of a SGW is not a frame
artefact. Consider a two-parameters congruence of geodesics. Let $\gamma_{1}(\tau,r)$
and $\gamma_{2}(\tau,r)$ be two neighboring geodesics with initially-parallel
tangent vectors $u^{\alpha}$ and connecting vectors $n^{\beta}.$

The selector parameter $r$ tells \char`\"{}which\char`\"{} geodesic
is being considered, while the affine parameter $\tau$ tells \char`\"{}where\char`\"{}
on a given geodesic someone is. Then, the relative acceleration of
these two neighboring geodesics is given, at the lowest order of approximation
\cite{key-23,key-26}, by the Jacobi-Levi-Civita (JLC equation) of
geodesic spread, 

\begin{equation}
\frac{D^{2}n^{\delta}}{d\tau^{2}}=R_{\alpha\beta\gamma}^{\quad\delta}u^{\alpha}n^{\beta}u^{\gamma}\label{eq: def geo}\end{equation}
where 

\begin{equation}
u^{\alpha}(\tau,r)=\frac{\partial x^{\alpha}(\tau,r)}{\partial\tau}\mid_{r=const}\label{eq: u}\end{equation}
is the unit vector tangent to the geodesic $\gamma(\tau,r)$ and

\begin{equation}
n^{\alpha}(\tau,r)=\frac{\partial x^{\alpha}(\tau,r)}{\partial r}\mid_{\tau=const}\delta r\label{eq: n}\end{equation}
is the separation vector between two nearby geodesics. The quantity
$R_{\alpha\beta\gamma}^{\quad\delta}$ in (\ref{eq: def geo}) is
the Riemann curvature tensor defined in the standard way as $R_{\alpha\beta\gamma}^{\quad\delta}=\partial_{\beta}\Gamma_{\alpha\gamma}^{\delta}-\partial_{\gamma}\Gamma_{\alpha\beta}^{\delta}+\Gamma_{\sigma\beta}^{\delta}\Gamma_{\alpha\gamma}^{\sigma}-\Gamma_{\sigma\gamma}^{\delta}\Gamma_{\alpha\beta}^{\sigma}$
and is calculated along the central geodesic. The Christoffel connection
coefficients $\Gamma_{\mu\nu}^{\alpha}$ are defined by $\Gamma_{\mu\nu}^{\alpha}=\frac{1}{2}g^{\alpha\sigma}(\partial_{\mu}g_{\sigma\nu}+\partial_{\nu}g_{\mu\sigma}-\partial_{\sigma}g_{\mu\nu}).$
The operator $\frac{D}{d\mu}$ is the covariant derivative calculated
along that line. It is assumed that $r=0$ corresponds to the central
geodesic line while a second nearby geodesic corresponds to $r=r_{0}$.
To discuss the \emph{magnetic} component of motion in the field of
a SGW we require the geodesic deviation equations be extended to the
next order approximation. These equations were obtained in \cite{key-27},
while a modified derivation can be found in \cite{key-28}. Consider
the vector $w^{\alpha}$ defined as 

\begin{equation}
w^{\alpha}=\frac{Dn^{\alpha}}{dr}=n_{;\beta}^{\alpha}n^{\beta}=\frac{\partial^{2}x^{\alpha}}{\partial r^{2}}+\Gamma_{\beta\gamma}^{\alpha}u^{\beta}u^{\gamma}.\label{eq: w}\end{equation}
This vector obeys the equations \cite{key-3,key-27,key-28}\begin{equation}
\frac{D^{2}w^{\delta}}{d\tau^{2}}=R_{\alpha\beta\gamma}^{\quad\delta}u^{\alpha}u^{\gamma}w^{\beta}+(R_{\alpha\beta\gamma;\epsilon}^{\quad\delta}-R_{\gamma\epsilon\alpha;\beta}^{\quad\delta})u^{\alpha}u^{\beta}u^{\gamma}u^{\epsilon}+4R_{\alpha\beta\gamma}^{\quad\delta}u^{\beta}\frac{Dn^{\alpha}}{d\tau}n^{\gamma}.\label{eq: geo estese}\end{equation}
Defining the vector

\begin{equation}
N^{\alpha}\equiv r_{0}n^{\alpha}+\frac{1}{2}r_{0}^{2}w^{\alpha},\label{eq: N}\end{equation}
equations (\ref{eq: def geo}) and (\ref{eq: w}) can be combined
to give \cite{key-3,key-27,key-28}\begin{equation}
\frac{D^{2}N^{\delta}}{d\tau^{2}}=R_{\alpha\beta\gamma}^{\quad\delta}u^{\alpha}u^{\gamma}N^{\beta}+(R_{\alpha\beta\gamma;\epsilon}^{\quad\delta}-R_{\gamma\epsilon\alpha;\beta}^{\quad\delta})u^{\alpha}u^{\beta}N^{\gamma}N^{\epsilon}+2R_{\alpha\beta\gamma}^{\quad\delta}u^{\beta}\frac{DN^{\alpha}}{d\tau}N^{\gamma}+O(r_{0}^{3}).\label{eq: geo estesissime}\end{equation}
It is then possible to write the expansion of $x^{\alpha}(\tau,r_{0})$
in terms of $N^{\alpha}$ as

\begin{equation}
x^{\alpha}(\tau,r_{0})=x^{\alpha}(\tau,0)+N^{\alpha}-\Gamma_{\beta\gamma}^{\alpha}N^{\beta}N^{\gamma}+O(r_{0}^{3}).\label{eq: exp}\end{equation}
This formula shows that in the frame of the local observer (in which
the Christoffel connection coefficients $\Gamma_{\beta\gamma}^{\alpha}=0$
along the central geodesic line \cite{key-23}) the spatial components
of N will directly give the time-dependent position of the nearby
test mass. According to (\ref{eq: geo estesissime}), these positions
include the next-order corrections as compared with solutions to (\ref{eq: def geo}).
We now specialize to the SGW metric (\ref{eq: metrica TT scalare}),
and take into account only linear perturbations in the SGW amplitude.
The first test mass is described by the central timelike geodesic
$x^{i}(t)=0$, its tangent vector being $u^{\alpha}\equiv(1,0,0,0)$.
The second test mass is situated at the unperturbed position $x^{i}(0)=l^{i},$
having zero unperturbed velocity. Assuming that the frame of the local
observer is located along the central geodesic, the task is to find
the trajectory of the second test mass using the generalized geodesic
deviation equation (\ref{eq: geo estesissime}). The deviation vector
can be written as 

\begin{equation}
N^{i}(t)=l^{i}+\delta l^{i}(t)\label{eq: N2}\end{equation}
where the variation in distance $\delta l^{i}(t)$ is caused by the
SGW. In the frame of the local observer one can replace all covariant
derivative in (\ref{eq: geo estesissime}) by ordinary derivatives
\cite{key-23}. In the lowest approximation (\ref{eq: geo estesissime})
reduces to (\ref{eq: def geo}) and specializes to

\begin{equation}
\frac{d^{2}\delta l^{i}(t)}{dt^{2}}=-\frac{1}{2}l^{j}\frac{\partial^{2}}{\delta t^{2}}\Phi\delta_{j}^{i}=\frac{1}{2}\omega^{2}l^{j}\Phi e_{\textrm{ $j$ }}^{(s)i}\label{eq: delta 1}\end{equation}
in the field of (\ref{eq: metrica TT scalare}). The relevant solution
to (\ref{eq: delta 1}) coincides exactly with the usual \emph{electric}
part of the motion given by (\ref{eq: traditional}). Since we want
to identify the \emph{magnetic} part of the gravitational force arising
from a SGW, all terms in (\ref{eq: geo estesissime}) must be considered.
Since $\frac{DN^{a}}{d\tau}$ is of order $\Phi$, the third term
of (\ref{eq: geo estesissime}) is of order $\Phi^{2}$ and can be
neglected. Computing the derivatives of the curvature tensor and substituting
them into (\ref{eq: geo estesissime}) specialized in the field of
(\ref{eq: metrica TT scalare}), the correct equations of motion read:\begin{equation}
\frac{d^{2}\delta l^{i}(t)}{dt^{2}}=\frac{1}{2}\omega^{2}l^{j}\Phi e_{\textrm{ $j$ }}^{(s)i}-\frac{1}{2}\omega^{2}l^{k}l^{l}(k_{j}\delta^{ij}+\frac{1}{2}k^{i})\delta_{l}^{j}\Phi e_{kj}^{(s)}.\label{eq: delta 2}\end{equation}
The second term on the right hand side of (\ref{eq: delta 2}) is
responsible for the \emph{magnetic} component of motion and can be
interpreted as the gravitational analogue of the magnetic part of
the Lorentz force (see also the analogy for ordinary tensorial waves
in \cite{key-3} and \cite{key-4}).

\section{Variation of distances between test masses and response of interferometers }

$\mbox{ }\mbox{ \mbox{ \mbox{ }}}$ In this Section we are interested
in the distance between the central particle, located at the coordinate
origin and a particle located on average at some position $(l_{1},l_{2},l_{3})$.
This model represents the situation of the beam-splitter and an interferometer
\cite{key-3,key-4}. In this frame, the Galilean distance - accurate
to terms of order $\Phi l$ and $\Phi l^{2}\omega$ while neglecting
terms quadratic in $\Phi$ - is given by, \begin{equation}
d(t)=\sqrt{x^{2}+y^{2}+z^{2}}+O((\Phi l(\omega l)){}^{2}),\label{eq: metrica local lorentz 2}\end{equation}
Letting \begin{equation}
\begin{array}{ccc}
x=l_{1}+\delta x, & y=l_{2}+\delta y, & z=l_{3}+\delta z,\end{array}\label{eq: delta xyz}\end{equation}
in (\ref{eq: metrica local lorentz 2}) we get\begin{equation}
d(t)=l+\frac{1}{l}(l_{1}\delta x+l_{2}\delta y+l_{3}\delta z).\label{eq: distanza}\end{equation}
By using the time dependent positions (\ref{eq: Grishuk 01}), the
distance $d(t)$ takes the form (with the required approximation $\omega l\ll1$):\begin{equation}
d(t)=l+\frac{1}{2l}(l_{1}^{2}-l_{2}^{2})\Phi(\omega t)-\frac{1}{4l}\omega l_{3}(l_{1}^{2}-l_{2}^{2})\Phi(\omega t-\frac{\pi}{2}).\label{eq: distanza 2}\end{equation}

The first correction to $l$ is due to the \emph{electric} contribution,
while the second correction to $l$ is due to the \emph{magnetic}
contribution. According to (\ref{eq: Grishuk 01}), the \emph{magnetic}
component of motion is present even if the mean position of the second
test mass is such that $l_{3}=0$, in perfect analogy with the pure
General Relativity case shown in \cite{key-3,key-4}.

Now we consider the response of a laser interferometer. To compute
the response function of the interferometer to a SGW from arbitrary
propagating directions we recall that the arms of the interferometer
are in the $\vec{u}$ and $\vec{v}$ directions, while the $x,y,z$
frame is adapted to the propagating SGW. Therefore, performing a spatial
rotation of the coordinate system, we obtain 

\begin{equation}
\begin{array}{ccc}
u & = & -x\cos\theta\cos\phi+y\sin\phi+z\sin\theta\cos\phi\\
\\v & = & -x\cos\theta\sin\phi-y\cos\phi+z\sin\theta\sin\phi\\
\\w & = & x\sin\theta+z\cos\theta,\end{array}\label{eq: rotazione}\end{equation}
or, in terms of the $x,y,z$ frame:

\begin{equation}
\begin{array}{ccc}
x & = & -u\cos\theta\cos\phi-v\cos\theta\sin\phi+w\sin\theta\\
\\y & = & u\sin\phi-v\cos\phi\\
\\z & = & u\sin\theta\cos\phi+v\sin\theta\sin\phi+w\cos\theta.\end{array}\label{eq: rotazione 2}\end{equation}
In this way the SGW is propagating from an arbitrary direction $\vec{r}$
to the interferometer (see figure 2). %
\begin{figure}
\includegraphics{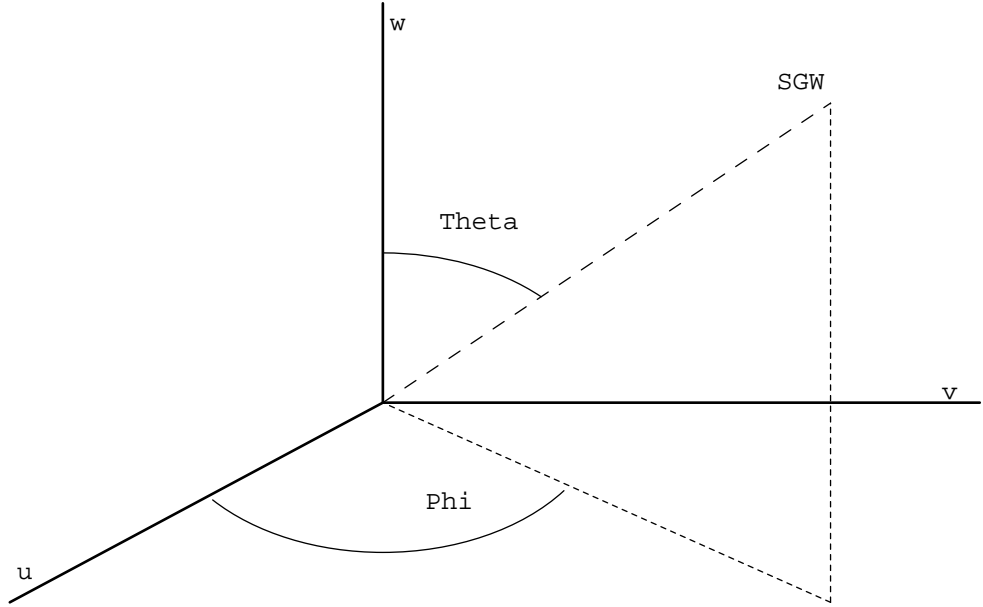}

\caption{A SGW propagating from an arbitrary direction}

\end{figure}
The response function $\delta d(t)$ is defined by

\begin{equation}
\delta d(t)\equiv d_{u}(t)-d_{v}(t),\label{eq: funzione risposta}\end{equation}
where $d_{u}(t)$ and $d_{v}(t)$ are the distances in the $u$ and
$v$ direction. Using equations (\ref{eq: distanza 2}), (\ref{eq: rotazione}),
(\ref{eq: rotazione 2}) and (\ref{eq: funzione risposta}) we obtain

\begin{equation}
\delta d(t)=-\Phi(t)l\sin^{2}\theta\cos2\phi+\frac{1}{4}\Phi(t)\omega l^{2}\cos\theta\left\{ \left[(\frac{1+\sin^{2}\theta}{2})+\sin^{2}\theta\sin2\phi\right](\cos\phi-\sin\phi)\right\} .\label{eq: funzione risposta 2}\end{equation}

The first term in (\ref{eq: funzione risposta 2}) is due to the \emph{electric}
contribution, while the second term is due to the \emph{magnetic}
contribution. Our response function (\ref{eq: funzione risposta 2})
is more complete than the previously derived expression in \cite{key-19,key-21,key-29},
since it includes an intrinsic \emph{magnetic} contribution. The function
$\frac{1}{4}\omega l\cos\theta[((\frac{1+\sin^{2}\theta}{2})+\sin^{2}\theta\sin2\phi)(\cos\phi-\sin\phi)]$
represents the so-called {}``angular pattern'' \cite{key-19} of
interferometers for the \emph{magnetic} contibution of SGWs. A plot
of the absolute value of this function is shown for the Virgo and
LIGO interferometers in Figure 3 and 4 respectively, for a frequency
of $f=1000Hz$ in each case. %
\begin{figure}
\includegraphics{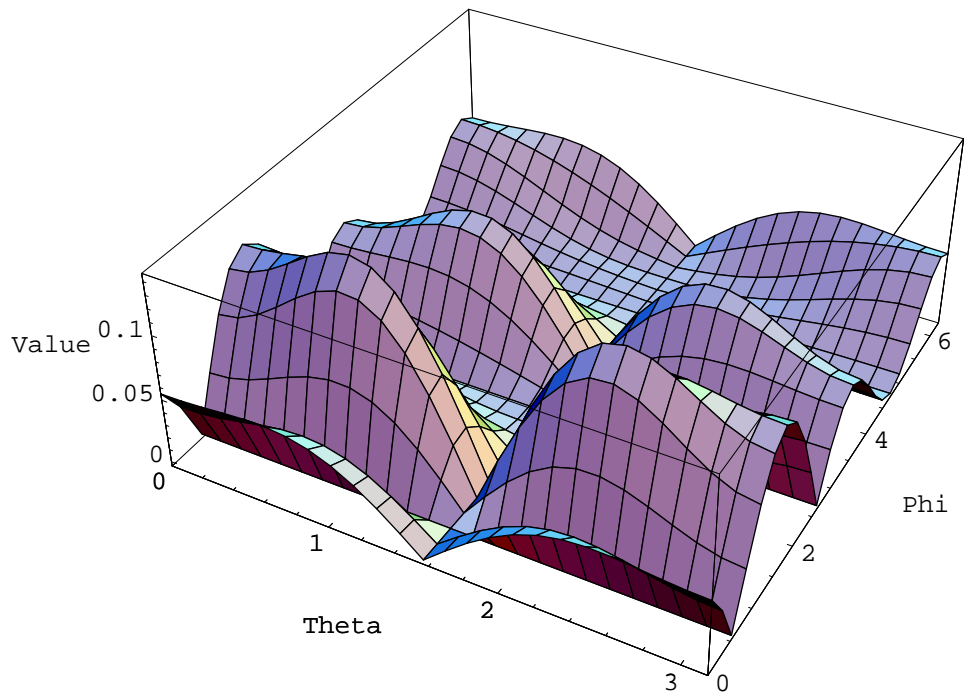}

\caption{The absolute value of the angular dependence of the total response
function of the Virgo interferometer to the magnetic component of
a SGW for $f=1000Hz$}

\end{figure}
\begin{figure}
\includegraphics{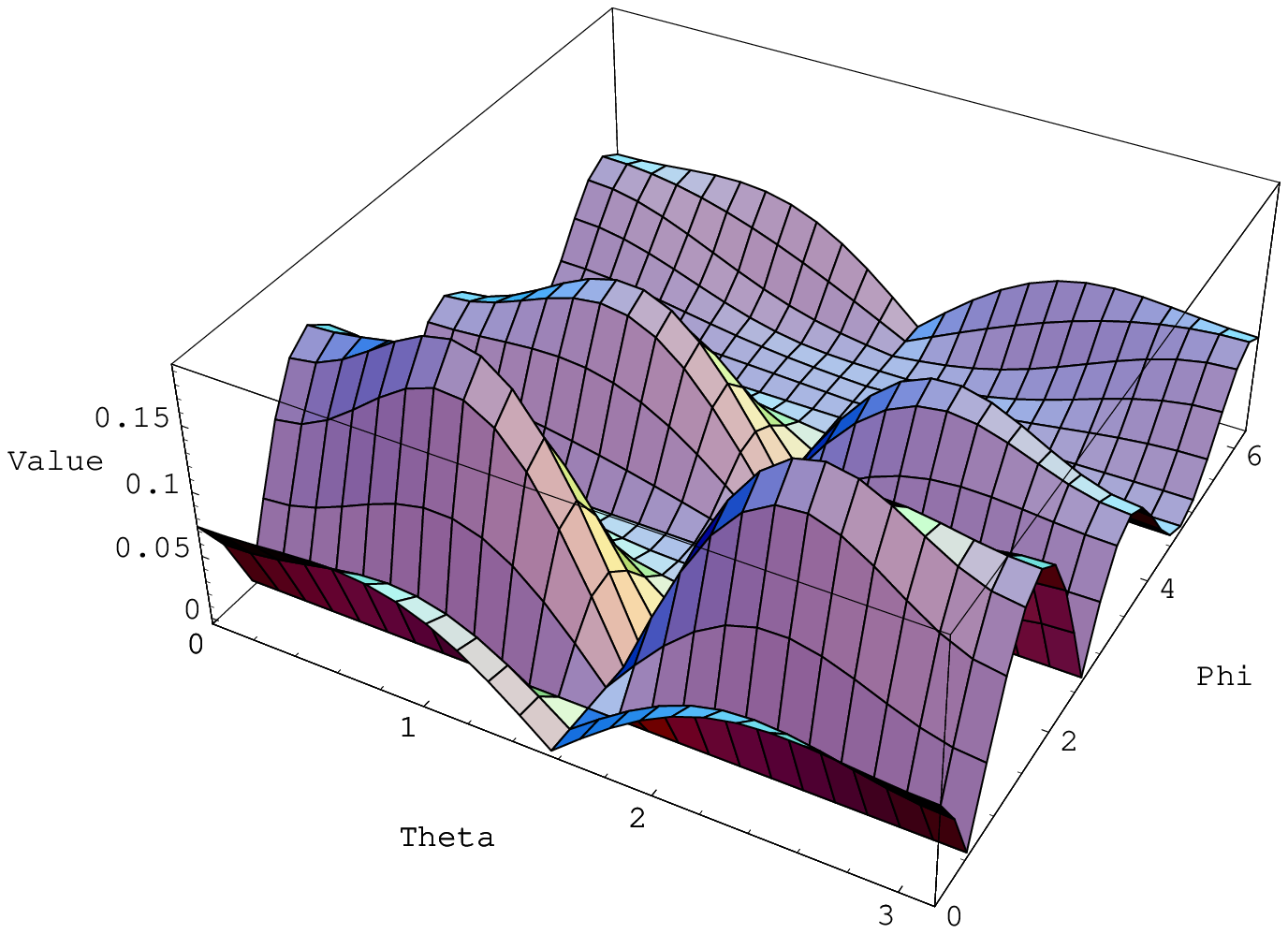}

\caption{The absolute value of the angular dependence of the total response
function of the LIGO interferometer to the magnetic component of a
SGW for $f=1000Hz$}

\end{figure}

\section{Variation of distance from the theory of Scalar Gravitational Waves}

$\mbox{ }\mbox{ \mbox{ \mbox{ }}}$ It is important to verify that
the approximate expression (\ref{eq: distanza 2}) follows directly
from the exact definition of distance which arises from the theory
of SGWs. For this purpose we make use of the time of flight measurement
of a photon travelling from one test mass to another and back. Indeed,
such consideration of photon transit time is an integral part of the
more general problem of finding the null geodesics of light in the
presence of weak gravitational waves \cite{key-3,key-4,key-19,key-20,key-23,key-24,key-25}.
This definition of distance $d(t)$ is valid independent of the relationship
between $l$ and $\omega$ and does not require the introduction of
the frame of the local observer. Consider a photon emitted from the
beam-splitter of an interferometer at instant $t_{0}$. Furthermore,
assume the same photon is reflected by a mirror and returns to the
beam-splitter at instant $t_{1}$. The proper distance $d(t)$ between
the two test masses at time $t$ is defined as \begin{equation}
d(t)=\frac{t_{1}-t_{0}}{2},\label{eq: distanza 3}\end{equation}
where $t=\frac{t_{1}+t_{0}}{2}$ is the mean time between the departure
and arrival of the photon. In the field of a SGW, the trajectories
of photons are described by the null geodesic $ds^{2}=0,$ which from
(\ref{eq: metrica TT scalare}), leads to\begin{equation}
dt^{2}=(1+\Phi)dx^{2}+(1+\Phi)dy^{2}+dz^{2}.\label{eq: null geodesic}\end{equation}
The unperturbed outgoing photon can be parametrized as \cite{key-3}

\begin{equation}
\begin{array}{ccccccc}
t=t_{0}+l\zeta, &  & x=l_{1}\zeta, &  & y=l_{2}\zeta, &  & z=l_{3}\zeta,\end{array}\label{eq: tempi}\end{equation}
with $0\leq\zeta\leq1$. Thus, according to equation (\ref{eq: null geodesic})
we obtain

\begin{equation}
dt=l[1+\Phi(\zeta)\frac{l_{1}^{2}+l_{2}^{2}}{l^{2}}]d\zeta.\label{eq: null geodesic 2}\end{equation}
Integrating both the sides of this equation, we find the time of arrival
of the photon at the mirror. In an analogous manner, the unperturbed
reflected photon can be parametrized as \begin{equation}
\begin{array}{ccccccc}
t=t_{1}-t_{0}+l\zeta, &  & x=l_{1}-l_{1}\zeta, &  & y=l_{2}-l_{2}\zeta, &  & z=l_{3}-l_{3}\zeta,\end{array}\label{eq: tempi 2}\end{equation}
where it is $0\leq\zeta\leq1$ again. An analogous integration of
(\ref{eq: null geodesic 2}) gives the time of arrival of the photon
back at the beam splitter (from the mirror). Combining the two pieces
of the transit of the photon from the beam splitter to mirror and
back enables one to obtain an exact expression for the distance $d(t),$
namely 

\begin{equation}
d(t)=l+\frac{l_{1}^{2}-l_{2}^{2}}{4l}\left[\frac{\Phi(\omega t+\omega l_{3})-\Phi(\omega t+\omega l)}{\omega(l-l_{3})}-\frac{\Phi(\omega t+\omega l_{3})-\Phi(\omega t-\omega l)}{\omega(l+l_{3})}\right].\label{eq: distanza perturbata}\end{equation}

In the low frequency approximation and retaining only the first two
terms in the expansion of $\Phi,$ (\ref{eq: distanza perturbata})
reduces to (\ref{eq: distanza 2}). Equation (\ref{eq: distanza 2})
is sufficient only for ground based interferometers, for which the
condition $\omega\ll1/l$ is in general satisfied. In the case of
space-based interferometers for which the above condition is not satisfied
in the high-frequency region of the sensitivity band \cite{key-3,key-4},
the theory of scalar gravitational waves (\ref{eq: distanza perturbata})
must be employed. Having shown that formula (\ref{eq: distanza 2})
follows from the exact distance (\ref{eq: distanza perturbata}) suggests
the \emph{magnetic} contribution to the distance is an universal physical
phenomenon which has to be taken into account in the data analysis
of GW interferometers.

\section{Conclusions}

$\mbox{ }\mbox{ \mbox{ \mbox{ }}}$ In this paper the motion of a
free test particle in the field of a scalar gravitational wave arising
from scalar-tensor theories of gravity is considered in detail by
solving higher order geodesic deviation equations. The presence and
significance of the so-called \emph{electric} and \emph{magnetic}
contributions to the distance function between interferometer test
masses have been shown. The angular dependence of the response function
of interferometers due to the \emph{magnetic} correction to the distance
function in the low-frequency approximation was obtained. These results
generalize previous works in the literature, where it has been shown
that the \emph{electric} and \emph{magnetic} contributions are unambiguous
in the long-wavelength approximation for ordinary GWs arising from
General Relativity. Finally, for arbitrary frequency range, the proper
distance between the two interferometer test masses is calculated
and its usefulness in the high-frequency limit for space based interferometers
is briefly considered. 

The response function (i.e., variation of distance) which has been
obtained is more complete than the previous derived expressions in
the literature (see \cite{key-16,key-19,key-21,key-30} for example),
because such a response function includes the \emph{magnetic} contribution.
In the high-frequency regime the division between \emph{electric}
and \emph{magnetic} components becomes ambiguous, thus requiring use
of the theory of scalar gravitational waves without (low-frequency)
approximation. The exact expression for the response function was
derived from which it was verified that the \emph{electric} and \emph{magnetic}
corrections follow from a series expansion of the exact result. We
also emphasize that the presence of the higher order terms in the
frequency dependent response function become crucial for space-based
interferometers, since the (low-frequency) condition $\omega\ll1/l$
is not satisfied in the high-frequency portion of the sensitivity
band of the interferometer. 

The response function contains contributions from two terms, namely
the \emph{electric} and \emph{magnetic}, both of which have been explicitly
derived and explained. 

The detector response function of laser gravitational wave interferometers
with arms of finite-size (few kilometers length) is commonly described
in terms of the long-wavelength approximation. In \cite{key-3}, it
was estimated that errors arising from the description of the detector
response in terms of such approximation can be as large as $10\%$.
In order to improve the estimation of parameters of gravitational
waves it was suggested that the linear-frequency approximation be
used, where the response function becomes the sum of two terms, namely
the electric one (frequency-independent term) and the magnetic one
(linear in the frequency). A detailed analysis of such theoretical
estimates as well as new estimates of high-frequency corrections to
the detector response appear in \cite{key-31}. In that work, it was
argued that errors arising from the description of the detector response
in the low-frequency approximation for standard General Relativity,
periodic GWs and isotropic stochastic background searches can be as
large as $1-2\%$ at $f=1.2kHz$. Moreover, the exact formula for
the detector response is recommended whenever the accuracy of the
detector response is important, especially in high-frequency regimes
($f=37.5kHz$ for LIGO interferometers). 

Our findings suggest the magnetic component of SGWs is an universal
physical phenomenon that must be taken into account so as not to discard
potentially relevant gravitational wave interferometer data.

\section{Acknowledgements}

We thank Salvatore Capozziello, Mauro Francaviglia, Maria Felicia
De Laurentis and Giancarlo Cella for helpful advice during this work.
We thank an anonymous Referee for useful comments which led to the
improvement of this article. Thanks is also extended to the European
Gravitational Observatory (EGO) consortium and to the Associazione
Scientifica Galileo Galilei for use of their computing facilities.

\end{document}